\documentclass[twoside]{dis09}
\usepackage[latin1]{inputenc}
\usepackage[dvips]{graphicx,epsfig,color}
\usepackage{wrapfig,rotating}
\usepackage{amssymb,amsmath,array}

\begin{document}
\title{Medium-modified fragmentation functions}

%***********************************************************************
% AUTHORS INFORMATION AREA
%***********************************************************************
\author{R. Sassot$^1$ M. Stratmann$^2$ and P. Zurita$^1$
%
% Optional short acknowledgment: remove next line if non-needed
\thanks{Partially supported by CONICET, ANPCYT, UBACYT, BMBF, and the Helmholtz Foundation.}
%
% DO NOT MODIFY THE FOLLOWING '\vspace' ARGUMENT
\vspace{.3cm}\\
%
% Addresses and institutions (remove "1- " in case of a single institution)
1- Departamento de Fisica, FCEN, Universidad de Buenos Aires\\
Pabellon 1, Ciudad Universitaria (1428) Buenos Aires, Argentina  
%
% Remove the next three lines in case of a single institution
\vspace{.1cm}\\
2- Institut f\"{u}r Theoretische Physik,
Universit\"{a}t Regensburg, 93040 Regensburg, Germany \\
Institut f\"{u}r Theoretische Physik und Astrophysik, Universit\"{a}t W\"{u}rzburg,
97074 W\"{u}rzburg, Germany\\
}
%***********************************************************************
% END OF AUTHORS INFORMATION AREA
%***********************************************************************

\maketitle

\begin{abstract}
We discuss preliminary results on  medium-modified fragmentation functions 
obtained in a combined NLO fit to data on semi-inclusive deep inelastic scattering 
off nuclei and hadroproduction in deuteron-gold collisions. 
\end{abstract}

\section{Motivation}

In the last few years there has been a significant improvement in the perturbative QCD 
description of hadroproduction processes, and more specifically, in the precise determination 
of fragmentation functions \cite{DSS}. One of the most interesting features of these extractions is that they 
not only reproduce the usual data on electron-positron annihilation into hadrons, but they describe 
as well other processes like semi-inclusive deep inelastic scattering (SIDIS) 
and hadroproduction in 
proton-proton collisions, with remarkable precision. This demonstrates the ideas of factorization 
and universality, which are the starting points for the perturbative QCD description, in the 
kinematical domain accessed by these experiments. 

In spite of this apparently successful scheme, fragmentation functions (FFs) are far from giving 
us the whole picture of hadronization; they are an efficient way to encode the non-perturbative 
information required for computing cross sections, but they don't tell us much about, for instance, 
the space-time evolution of the hadron formation, the origin of the differences between the 
different species, and many other non perturbative key issues. 

To have an insight into  these features, we need to attack the problem from a different angle, 
that could be for example, studying how hadronization proceeds in a nuclear environment.
In recent years there has been an increasing 
number of experiments pointing in this direction, exploring hadroproduction
off nuclear targets with mass $A$, such as very precise semi-inclusive eA 
measurements by HERMES \cite{HERMES}, as
well as experiments using dAu collisions at the RHIC \cite{PHENIX,STAR}. Both processes show clear signals
of non trivial nuclear dependence in the hadronization mechanism, that needs to be understood. 
Additionally, this information serves as baseline for heavy ion research programs ongoing  
at RHIC and projected for LHC, aiming at unravelling the properties of hot and dense QCD matter \cite{TALKS}.

In parallel with the availability of data, several  theoretical calculations and models 
designed to describe the nuclear dependence have been proposed, some of them putting the 
emphasis on the interactions of hadrons and ``pre-hadrons'' in the nuclear medium, others focussing 
on the interactions of the ``seed partons'' with the medium. More recently, it has also been suggested 
that FFs could obey different energy scale dependence because of the nuclear medium.  
Most models reproduce with different success features of the data, in spite of very different, 
even orthogonal, approaches and ingredients \cite{TALKS}.

 In our approach, rather than proposing a particular mechanism, in a first step we try to isolate 
 the medium induced modification of the FFs as precisely as possible, factoring out the standard 
 knowledge on nuclear effects in the parton densities, as seen in deep inelastic scattering or Drell 
 Yan processes with nuclei \cite{nDS}, and the QCD dynamics inherent to hard processes at NLO accuracy. 
 In other words, we fit medium-modified fragmentation functions (nFFs), assuming (or testing) 
 factorization and universality in a nuclear environment. 

\section{Convolutional Approach:} 

Rather than fitting from scratch the nFFs, which would take as many parameters as the standard or 
vacuum FFs, plus some more to represent the nuclear size or density dependence, in the following 
we choose to relate the nFFs to the standard ones by a convolution:
\begin{equation}
D^{h}_{i/A}(z,Q^2_0) =\int_z^1 \frac{dy}{y} W_i(y,A,Q^2_0) D^{h}_{i}(\frac{z}{y},Q^2_0), 
\end{equation} 
where the weight function $W_i(y,A,Q^2_0)$ parameterizes nuclear effects in FFs at a given initial 
scale $Q_0$. The scale dependence of nFFs is then determined by ordinary NLO evolution 
equations. A simple delta function $\delta(1-y)$ as weight would imply no nuclear effects, while a shift in its 
argument, i.e.\ $\delta(1-\epsilon-y)$, would represent a 
shift in the momentum fraction as suggested, for example, by some 
energy loss mechanism. A more general weight function like
\begin{equation}
W_i(y,A,Q^2_0)=n_i \,y^{\alpha_i}(1-y)^{\beta_i},
\end{equation}
can parameterize effects not necessarily related to modifications the parton's momentum, such as hadron 
or pre-hadron attenuation or enhancement, with a great economy of parameters and, at the same time, 
retaining the information on the vacuum FFs. The $A$ dependence of the weights can easily be 
included in the coefficients taking them as smooth functions of $A$. On the other hand, convolutional 
integrals are the most natural language for parton dynamics beyond the LO and allow the application 
of the Mellin transform technique in the NLO computation of the scale dependence and cross section estimates. 

The convolutional approach has been shown to be specially effective in the extraction of initial state 
nuclear effects in inclusive DIS and Drell Yan processes at NLO accuracy \cite{nDS}. For nFFs this is also
the case, and it can be shown that an extremely simple functional form for the weights with very few 
parameters can reproduce the main features of the available data. However, since we are interested 
in obtaining an accurate parameterization, we implement more flexible weights to account for 
mechanisms other than global shifts in momentum but still preserving factorization. Specifically, we
adopt the following ansatz:
\begin{eqnarray}
W_q(y,A,Q^2_0)=n_q\delta(1-\epsilon_q-y)+n'_q \, y^{\alpha_q} (1-y)^{\beta_q} \\
W_g(y,A,Q^2_0)=n_g\delta(1-\epsilon_g-y)+n'_g \, y^{\alpha_g} (1-y)^{\beta_g} 
\end{eqnarray}
discriminating between quarks and gluons. In order to enhance the sensitivity of the fit  to the gluon 
fragmentation, it is essential to include in the fit also hadroproduction rates from dAu collisions at 
RHIC \cite{PHENIX,STAR}. 
%Since the sensitivity of SIDIS data \cite{HERMES} to the gluon 
%fragmentation is rather limited, it is essential to include in the fit also hadroproduction rates from 
%dAu collisions at RHIC \cite{PHENIX,STAR}. 
The $A$ dependence of the coefficients is implemented as 
$n_i=\lambda_{n_i}+\gamma_{n_i}\,A^{\delta_{n_i}}$. In most cases, however, $\lambda$ can be set 
to zero or to unity due to the vanishing of nuclear effects as $A\to 1$. The assumption of a linear 
behavior ($\delta=1$) does not spoil the quality of the fit. 

Results of a fit with such an ansatz can be found in Fig.~1-3, for 
SIDIS pion multiplicities off different nuclei from HERMES \cite{HERMES} (rates are normalized to a deuterium target),
neutral and charged pion production in dAu collisions from PHENIX \cite{PHENIX} and
STAR \cite{STAR}, respectively.
\begin{wrapfigure}{r}{0.66\columnwidth}
\vspace*{-0.5cm}\centerline{\includegraphics[width=0.68\columnwidth]{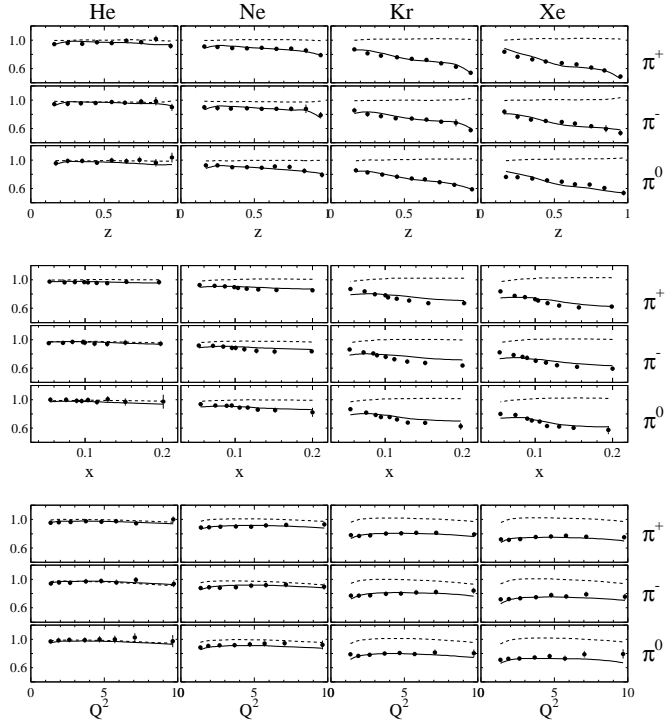}}
\vspace{-1cm}
\caption{Pion SIDIS multiplicities from HERMES }\label{Fig:pi}
\end{wrapfigure}
The dashed lines represent the estimates for these cross sections computed by taking into account
nuclear effects {\em only} for the PDFs. The solid lines refer to our results 
with nuclear modifications also in the FFs.
We obtain an overall $\chi^2=350.45$ for 368 data points and 
17 parameters, resulting in $\chi^2/d.o.f=0.997$. 
Focusing on SIDIS data in Fig.~1, the quality of the fit is quite impressive, and even the 
$x_{Bj}$ (or $\nu$) dependence is reproduced although the weights in Eqs.~3 and 4 and the nFFs only depend 
on $z$ as required in a factorized approach. 
In addition, there seems to be no conflict with the standard (vacuum) $Q^2$ dependence assumed in our fit. 
RHIC data are also reproduced well as shown in Figs~2 and 3. 

\begin{wrapfigure}{l}{0.5\columnwidth}
\vspace{-0.7cm}\centerline{\includegraphics[width=0.45\columnwidth]{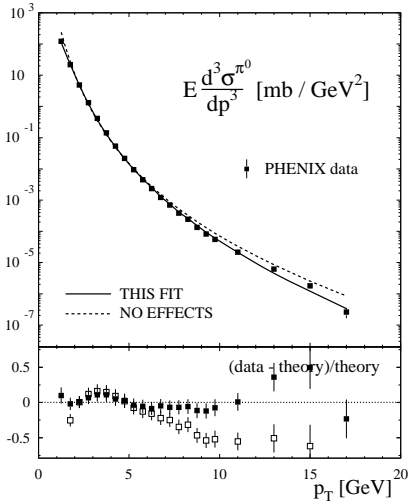}}
\vspace{-0.5cm}\caption{$\pi^0$ production in dAu collisions}\label{Fig:phenix}
\end{wrapfigure}
The logarithmic fall off of the cross section over four orders of magnitude hinders the 
comparison between data and the theoretical estimates with and without nFFs.
The differences are sizable, as shown in the lower panel of Fig.~2: 
filled and open squares are the results for ``(data-theory)/theory'' for fits with and 
without modified FFs, respectively.
PHENIX data indicate
a significant suppression at larger $p_T$, while STAR data only hint
at such a behavior towards the end of the more
limited $p_T$ range. Even though the fit follows the data down to very low $p_T$, this
needs to be taken with extreme caution since possible soft contributions may spoil factorization,
and standard PDFs and FFs do not give a satisfactory description of pp data at NLO below
$p_T\simeq 2\,\mathrm{GeV}$.

\begin{wrapfigure}{r}{0.66\columnwidth}
\vspace{-0.7cm}\centerline{\includegraphics[width=0.7\columnwidth]{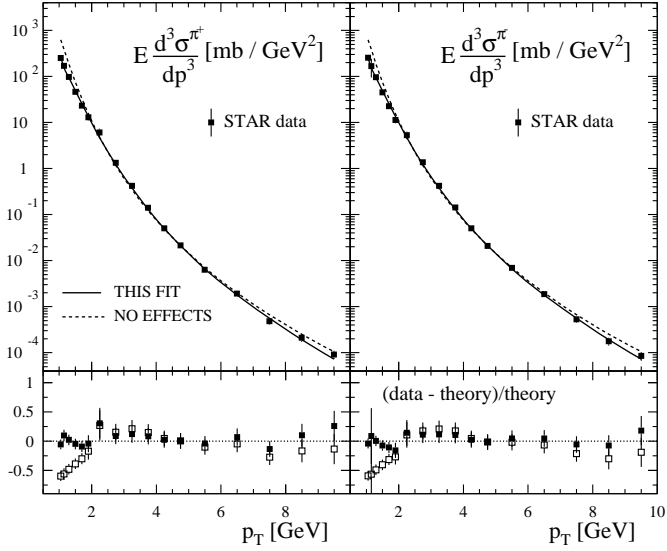}}
\vspace{-0.3cm}\caption{$\pi^{\pm}$ production in dAu collisions}\label{Fig:star}
\end{wrapfigure}
In Table 1 we present the values obtained for the 17 free parameters of the fit. Table 2 shows the
values the coefficients in the weight functions in Eqs.~3 and 4 take for different nuclei
using the parameters in Table 1. 
As it can be noticed, the $A$ dependence of the coefficients is rather smooth and typically linear 
for $A$ not too small. 
Quark and gluon coefficients are found to be rather different.  
In the case of quarks (and antiquarks) the main modification due to the nuclear medium is related
to the first term in Eq.~3, associated to a shift $\epsilon_q$ in the momentum fraction. This shift is 
typically small, but introduces a non trivial $z$ dependence, and increases linearly with $A$. 
The normalization of this term, in turn, decreases with $A$ leading to an overall suppression. More than
80\% of the nuclear modification comes from this term, while the second term represent a smaller effect.
For gluons, both the first and the second term compete resulting in a much more pronounced effect.

%\begin{wraptable}{l}{0.66\columnwidth}
\begin{table}[h!]
\begin{tabular}{|lcccccccccc|}\hline
                  &  $n_q$ & $\epsilon_q$ & $n'_q$& $\alpha_q$ & $\beta_q$ & $n_g$ &  $\epsilon_g$ & $n'_g$& $\alpha_g$ & $\beta_g$ \\ \hline
$\lambda$ &         1    &      0         &    0       &27.27&33.77&      1    &  0          & 0          &  16.05 &49.75 \\
$\gamma$ &  -0.045  & 0.00015  & 0.0016& -0.045& -1.932& 0.219& 0.0007&  -0.857& 0.021  & -0.128 \\
$\delta$     &   0.458  &       1         &    1          &  1          &     1      & 0.122&   1         & 0.078  &    1       &1     \\
\hline\end{tabular}
\caption{\label{tab:table1} Parameters describing the medium-modified FFs at NLO accuracy}
\end{table} 
%\end{wraptable} 

\begin{table}[h!]
\begin{tabular}{|lcccccccccc|}\hline
  $A$    &  $n_q$ & $\epsilon_q$ & $n'_q$& $\alpha_q$ & $\beta_q$ & $n_g$ &  $\epsilon_g$ & $n'_g$&
  $\alpha_g$ & $\beta_g$ \\ \hline
  He &    0.913  &  0.001  &  0.006  & 27.095 &  30.711 &  1.260 &   0.003  & -0.956 &  16.142 &  49.243 \\   
  Ne &    0.818  &  0.003  &  0.032  & 26.374 &  28.534 &  1.317 &   0.015  & -1.085 &  16.480 &  47.190 \\   
  Kr  &    0.649  &  0.013  &  0.134  & 23.490 &  25.317 &  1.377 &   0.063  &-1.215 &  17.830 &  38.979  \\  
  Ze &    0.570  &  0.020  &  0.210  & 21.372 & 23.967  &  1.398  &  0.098  & -1.258 &  18.822 &  32.949  \\  
  Au &    0.481  &  0.030  &  0.315  & 18.398 & 22.538  &  1.419  &  0.147  & -1.299 &  20.214 &  24.481  \\  
  \hline
\end{tabular}
\caption{\label{tab:table2}Coefficients of the weight functions in Eqs.~3 and 4 for different nuclei}
\end{table} 
It is also enlightening to analyze the ratios of the nuclear and vacuum 
fragmentation functions as a function of $z$ for a given $Q^2$, as shown in the left panels of Fig.~4. 
We also show the nFFs themselves in the right panels, and always at $Q^2=10\,\mathrm{GeV}^2$.

\begin{wrapfigure}{r}{0.66\columnwidth}
\vspace{-0.75cm}\centerline{\includegraphics[width=0.7\columnwidth]{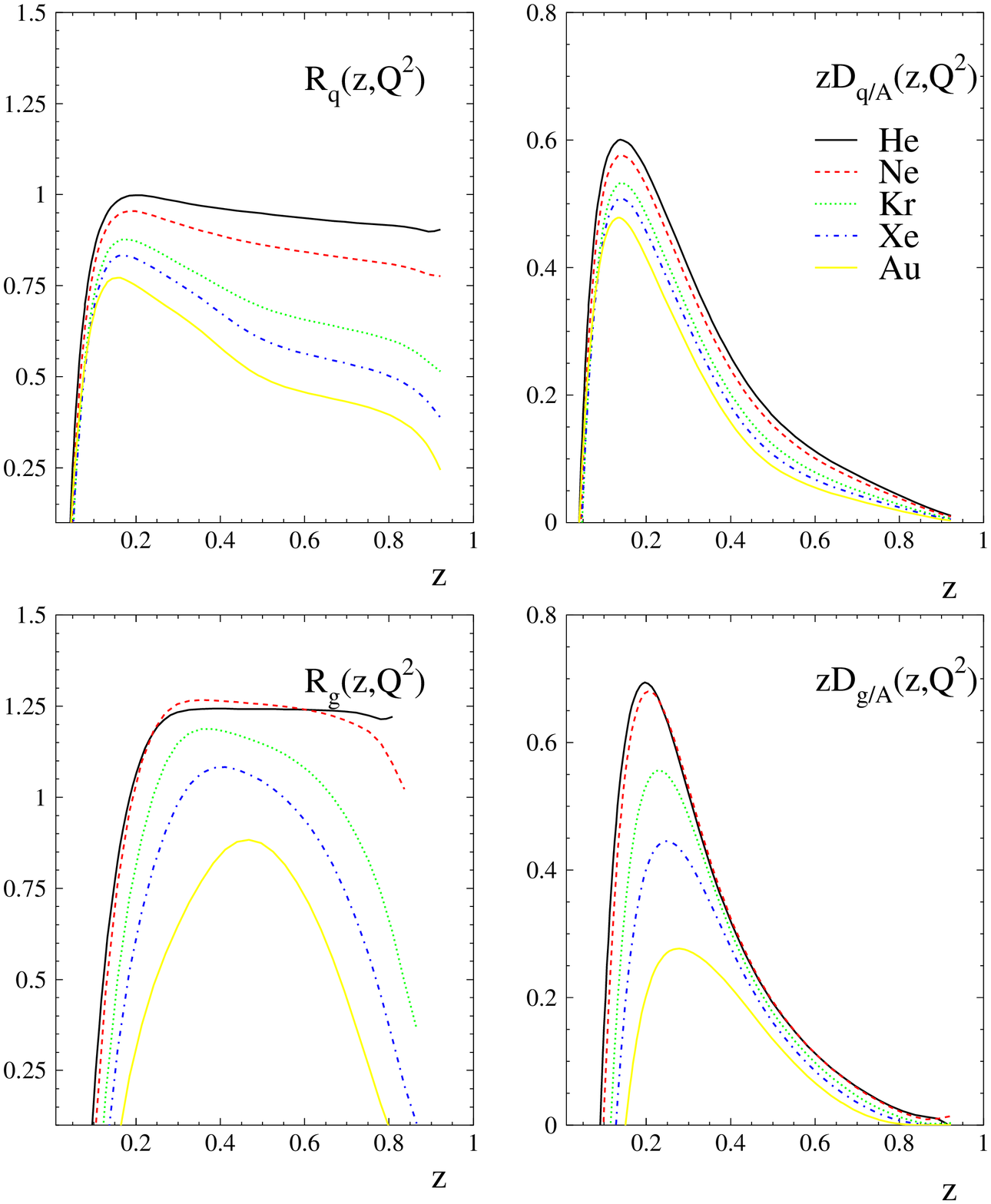}}
\vspace{-0.3cm}\caption{nFFs and rates at $Q^2=10\,\mathrm{GeV}^2$}\label{Fig:rates}
\end{wrapfigure}
For quarks and antiquarks the main effect seems to be a reduction 
in the fragmentation probability increasing with $z$ and with nuclear 
mass $A$. There is a significant drop at small $z$ ($z<0.1$), however, neither the nuclear nor the vacuum 
fragmentation functions are well constrained by the data, and neglected final state hadron mass effects can 
be significant. Gluon fragmentation shows a rather different pattern of medium modification, with a much more 
noticeable drop at large $z$, combined with a sizable enhancement  at intermediate $z$.  
Notice that the low $z$ behavior of the gluon nFFs is crucial for reproducing the $x_{Bj}$ dependence shown 
by the SIDIS rates and therefore it is relatively well constrained by these data. On the contrary, the large $z$
behavior is related to the suppression of the RHIC cross sections at low $p_T$ that need better understanding 
from the theoretical side.

\section{Conclusions}
We have explored the feasibility of reproducing the main features of data sensitive to medium induced 
modifications of fragmentation functions in a factorized QCD framework at NLO accuracy based on 
a convolutional approach.
The nice agreement found encourages us to further pursue this approach and extend it to hadrons
species other than pions.

% ****************************************************************************
% BIBLIOGRAPHY AREA
% ****************************************************************************

\begin{footnotesize}

\end{footnotesize}

% ****************************************************************************
% END OF BIBLIOGRAPHY AREA
% ****************************************************************************

\end{document}